\documentstyle[11pt,newpasp,twoside,epsfig]{article}
\input epsf.sty
\index{Pulsars!Radio}
\index{Pulsars!Emission Mechanism}
\index{Pulsars!Rotation}
\index{Pulsars!Aberration Retardation}
\reversemarginpar

\begin{document}
\title{Centrifugal Acceleration in Pulsar Magnetosphere }
 \author{R. M. C. Thomas and R. T. Gangadhara}
\affil{Indian Institute of Astrophysics, Bangalore -- 560034, India}
\begin{abstract}
We present a relativistic model of pulsar radio emission by plasma accelerated along
the rotating magnetic field lines projected on to a 2D plane perpendicular to the 
rotation axis. We have derived the expression for the trajectory of a particle, and 
estimated the spectrum of radio emission by the plasma bunches. We used the parameters
given in the paper by Peyman and Gangadhara (2002). Further the analytical expressions
for the Stokes parameters are obtained, and compared their values with the observed
profiles. The one sense of circular polarization, observed in many pulsars, can be
explained in the light of our model.
 \end{abstract}
\section{Introduction}
 It is important to understand the charged particle dynamics in the pulsar
magnetosphere  to unravel the radiation mechanism of pulsars.
The particles are constrained to move strictly along the field lines, owing to the 
super-strong magnetic field that the gyration of the particles are almost 
suppressed. The equation of motion for a charged particle moving along the rotating 
field line is given by Gangadhara (1996). Here we extend this work to obtain an 
analytical expression for particle trajectory and Stokes parameters. The pulsar 
rotation effects such as aberration and retardation can create asymmetric pulse profiles
  \section{Dynamics of a Charged Particle}
\label{sec:using}
The equation of motion of a charged particle along a rotating magnetic
field line is given by (Gangadhara 1996)
\begin{equation}
  \frac{d}{dt}\left (m\frac{d r}{dt}\right) =m {\Omega}^{*2} r,\quad\quad
 \gamma = \left (1- \frac{\dot r^2}{c^2}-
\frac{r^2\Omega^{*2}}{c^2}\right)^{-1/2}
  \end{equation}
where $ m=m_{0}{\gamma} $ is the relativistic electron mass, $m_0$ is the rest mass, c is the
speed of light,
$\gamma$  is the Lorentz factor, $\Omega^{*}=\Omega\sqrt{1-{b^2}/{r^2}}$ is the effective angular 
velocity of the particle, and $b = d_0 \cos\theta_0.$ We have solved this equation and obtained the 
analytical solution:
\begin{equation}
 r=\frac{c\sqrt{1+D^2}}{\Omega} cn(\lambda -\Omega t),
\end{equation}
where $\rm{cn}(\lambda-\Omega t)$ is the Jacobian Elliptical cosine function
      (Abramowitz $\&$ Stegun), 
$$ D= \frac{\Omega d_{0}\cos\theta_{0}}{c} ,          \quad 
   \lambda=\int\limits_{0}^{\phi_0}\frac{d\zeta}{\sqrt{1-k^2{\sin{\zeta}}^{2}
   }},\quad \phi_0=\arccos\left(\frac{r_0\Omega}{c}\right),$$
$r_o$ the initial particle injection point, $d_o$ the distance between 
magnetic pole and rotation axis in the projected 2D plane, $\theta_o$ the
initial injection angle with respect to  meridional plane, 
$\Omega$ the  angular velocity of the pulsar, k is an integration  constant, $\zeta$ is a dummy 
variable, $\theta$ is the angle that the line of sight makes with the 2D plane.
Using the expression for the 'r' an approximate value for $ \rho $, the radius 
of curvature of the particle in the lab frame, is found out to be $\rho\approx  \sqrt{1+D^2}\  
r_{L}/2 $ where $r_{L}$ is the light cylinder radius.
 
\begin{figure}
\plotone {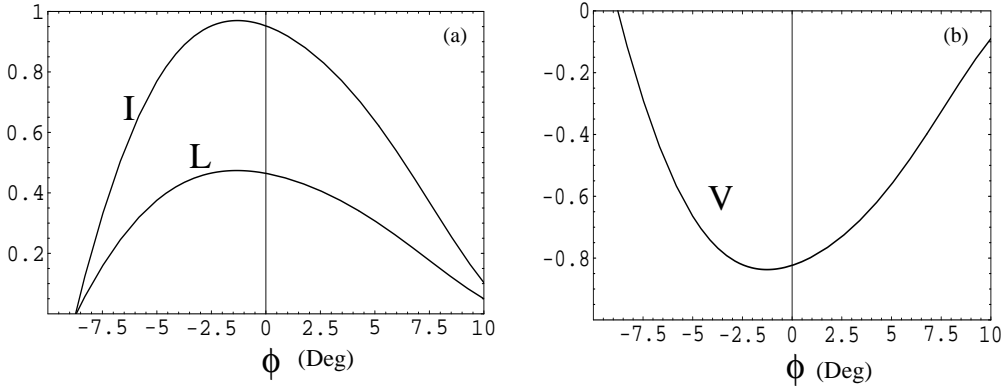}
\caption{\hskip -0.5 cm (a) Intensity and linear polarization vs rotation phase $\phi$ 
for the radiation emitted by particles accelerated along the rotating magnetic 
field lines, and (b) Circular polarization of the emitted radiation.} 
\end{figure}

For a pulsar with period of 1 s, we estimated the components of the
electric field of radiation and hence the Stokes parameters. In Fig.~1a, we
have plotted the intensity $I$, linear polarization $L,$ and in Fig.~1b
circular polarization $V$ is plotted. Due to aberration-retardation phase
shift the profile is asymmetric about the center $(\phi=0)$ of the profile.
It also shows that leading part of the profile is stronger than trailing part.
Since line of sight stays above or below the plane of particle trajectory, circular
polarization becomes one sense, as observed in many pulsars.

\end{document}